**SPORT: A new sub-nanosecond time-resolved instrument to study swift heavy ion-beam induced luminescence – Application to luminescence degradation of a fast plastic scintillator**

E. Gardés[a,b,*], E. Balanzat[a], B. Ban-d'Etat[a], A. Cassimi[a], F. Durantel[a], C. Grygiel[a],

T. Madi[a], I. Monnet[a], J.-M. Ramillon[a], F. Ropars[a], H. Lebius[a]

[a] CIMAP, CEA-CNRS-ENSICAEN-Université de Caen Basse Normandie,

Bd Henri Becquerel, BP 5133, 14070 Caen Cedex 5, France

[b] ISTO, CNRS-Université d'Orléans

1A Rue de la Férollerie, 45071 Orléans Cedex 02, France

*Corresponding author. E-mail address: emmanuel.gardes@cnrs-orleans.fr

**Abstract**

We developed a new sub-nanosecond time-resolved instrument to study the dynamics of UV-visible luminescence under high stopping power heavy ion irradiation. We applied our instrument, called SPORT, on a fast plastic scintillator (BC-400) irradiated with 27-MeV Ar ions having high mean electronic stopping power of 2.6 MeV/µm. As a consequence of increasing permanent radiation damages with increasing ion fluence, our investigations reveal a degradation of scintillation intensity together with, thanks to the time-resolved measurement, a decrease in the decay constant of the scintillator. This combination indicates that luminescence degradation processes by both dynamic and static quenching, the latter mechanism being predominant. Under such high density excitation, the scintillation deterioration of BC-400 is significantly enhanced compared to that observed in previous investigations, mainly performed using light ions. The observed non-linear behaviour implies that the dose at which luminescence starts deteriorating is not independent on particles' stopping power, thus illustrating that the radiation hardness of plastic scintillators can be strongly weakened under high excitation density in heavy ion environments.

.

**Introduction**

Scintillators play a central role in various fields including particle detection and discrimination, which is in constant evolution, or in emerging fields such as hadrontherapy, where heavy particles are involved. The knowledge of the yields and lifetimes as well as the radiation hardness of scintillators when, instead of electrons or X-rays, high excitation density particles are used is essential. Time-resolved measurement is a major asset to investigate the excitation and energy transfer processes involved in luminescence, which often occur over a few nanoseconds at most. To our knowledge, the only device allowing for the investigations of scintillators dynamics under swift heavy ion irradiation was developed by K. Kimura and collaborators [1-7]. With a resolution of about 85 ps, their instrument extended the field of investigations towards the primary processes under extremely high-density excitation. Especially, one of their remarkable outcome was to evidence a novel ultra-fast luminescence (UFL) in almost all of the insulators investigated [4-7]. UFL is a wide wavelength range emission that can be as short as several tens of picoseconds. This phenomenon cannot be attributed to known excited species but was interpreted to result from the relaxation of dense electron–hole plasma in the track core of swift heavy ions [4-7]. This is an example of the uniqueness of the phenomena induced by swift heavy ion irradiation, where energy deposition can be orders of magnitude higher than by electron or X-ray irradiation.

We present here the first version of a new device to study the dynamics of ion-beam induced luminescence (IBIL) using high stopping power heavy ion irradiation. Our instrument, called SPORT (French acronym for time-resolved optical spectroscopy), will allow for simultaneous UV-visible spectral and sub-nanosecond time-resolved measurements in a wide range of temperatures, thus enabling the discrimination of the different processes involved in scintillation and their evolution with irradiation. Our first investigations using SPORT were performed on a widely used plastic scintillator: BC-400, also known as NE-102A. Plastic scintillators cumulate many advantages such as fast and short time responses, facilitated shaping and low costs. Therefore they have been used for decades in many radiation applications [8]. Their main drawback is however the degradation of their luminescence properties under irradiation. It has long been known that the energy deposition in organic scintillators not only excites the fluorophores but also breaks chemical bonds, leading to luminescence quenching [9]. Accordingly, the previous IBIL studies of BC-400 revealed a degradation of scintillation as a function of irradiation dose [10-13]. However, these studies used light ions and were not time-resolved, as usually for studies of IBIL intensity degradation of plastic scintillators. We report here the first time-resolved IBIL investigations of BC-400 using swift heavy ion irradiation. Our work, focused on the dynamics of the scintillator and its evolution with ion fluence, brings new insights on the quenching type and the weakening of radiation hardness under high excitation.

**Experimental**

The instrumental set-up of SPORT is illustrated in Fig. 1. Basically, the principle of a time-resolved measurement with this instrument is to count photon in coincidence with ion hitting. Target samples are mounted in a UHV chamber on a cold finger tip (down to ~$10^{-9}$ mbar and 7.5 K). The sample holder is oriented normal to the ion beam and opens a 10-mm diameter window on the target. Photons are collected at 45° through two UV-vis windows and focused via spherical and planar mirrors on the entrance slits of two spectrographs (CP140-1604 and -1605 from HORIBA). They are finally counted using two 16-channel photomultiplier tubes (PMTs) disposed at the focal planes of the spectrographs (H10515B-04 and -20 from HAMAMATSU). This set-up allows to cover the 190-780 nm wavelength range with an average dispersion of ~14 nm/ch. Ions are detected with a home-made detector. Ions are passing through a 0.8 μm Al foil and the secondary electrons are detected by a tandem microchannel plate (MCP) (Fig. 1). After pre-amplification and constant fraction discrimination, the time-arrivals of the pulses from the PMTs and the ion detector are coded by a 32-channel multi-hit time-to-digital converter (TDC), which time resolution is adjustable down to 25 ps (V1290 from CAEN). Events (ion-photon coincidences) are sent to a computer and stored line by line, allowing for offline processing of the data. A detailed description of the final version of SPORT will be published elsewhere.

The sample investigated presently is a few mm-thick sheet of BC-400 from St-Gobain (equivalent to NE-102A). This ternary scintillator is composed of a polyvinyltoluene (PVT) matrix containing two fluorophores: p-Terphenyl (p-T) and 1,4-bis(5-phenyloxazol-2-yl) benzene (POPOP) [14]. The energy deposited by ionizing particles is transferred non-radiatively to p-T and then radiatively to POPOP. BC-400 has a maximum light output at 423 nm, which corresponds to the emission band of POPOP, and has rise and decay times of 0.9 and 2.4 ns respectively [14,15]. We set up SPORT in order to focus on the dynamics of this band and its evolution with ion fluence. Experiments were performed at the IRRSUD beam line of the Grand Accélérateur National d'Ions Lourds (GANIL) facility in Caen, France. The sample was irradiated at room temperature with 27-MeV $^{36}$Ar$^{10+}$ ions obtained from the injector cyclotron of GANIL facility operating in packages with a repetition rate of 13.45 MHz. According to SRIM code [16], these ions have a range of 11.3 μm and a mean electronic stopping power $<(dE/dx)_e>$ of 2.6 MeV/μm. This value is by far the highest among the previous IBIL studies of BC-400 [10-13]. The ion beam was collimated between 16×16 mm$^2$ slits onto the sample at a mean fluence rate of $3\times10^6$ ion cm$^{-2}$ s$^{-1}$. We used only one spectrograph and we present results on one channel of the PMT at 423 nm. Pulses from the PMT and the ion detector provided start and stop signals, respectively, on two channels of the TDC. The time resolution of the TDC was set to ~100 ps. The time-dependent emission curves of BC-400 were constructed by subtracting the time arrival of the photons to that of the ions and summing over the events of successive steps in fluence. As additional way to evaluate the time resolution of the instrument, and thanks to the multi-hit recording of the TDC, an ion self-correlation curve was constructed by subtracting the time arrival of consecutive stop signals, i.e. from the TDC channel provided by the ion detector.

**Results and discussion**

The ion self-correlation curve at final fluence, which represents the time spread between ions from successive packages of the cyclotron, is illustrated in Fig. 2. It is well described by a Gaussian function with a 0.6 ns-FWHM. The variance of this time spread represents twice the variance on the transit time from ion arrival to TDC time coding via ion detection and electronics. Therefore, any of the time jitters from this transit, for instance that of electronics, has to be less than $0.6/\sqrt{2} \approx 0.4$ ns.

The emission curve of the 423 nm band of BC-400 with 27-MeV Ar irradiation at final fluence is illustrated in Fig. 3. The rise part of the curve, defined from 10 to 90 % of pulse height, spreads over 1.3 ns, which is higher than the 0.9 ns usually reported for this scintillator [15]. Assuming that this difference completely originates from time jitter, we can estimate the upper bound for the total resolution to be $\sqrt{1.3^2 - 0.9^2} \approx 0.9$ ns. Therefore, the upper bound for the time jitter of the PMT can be calculated as $\sqrt{0.9^2 - 0.4^2} \approx 0.8$ ns. This indicates that the PMT is the component that limits the time resolution of our instrument.

As illustrated in Fig. 3, the final part of the pulse decay is well adjusted using a single exponential ($\propto e^{-t/\tau}$) where $\tau$ is the effective decay constant without correction of the resolution of the instrument. The fits of the emission curves acquired during successive fluence steps reveal that the decay is enhanced with irradiation (Fig. 4). The decay constant at initial fluence $\tau_0$ is 2.5 ns, which is in agreement with literature data for bulk BC-400 [15]. It decreases by about 20 % to reach 2 ns at the final fluence (Fig. 4). The intensity of light emission $I$ degrades even faster. It drops to less than 1/3 of its initial value ($I_0$) at the final fluence (Fig. 4).

As demonstrated by the brownish coloration of the sample surface after irradiation, it is obvious that both the decrease in decay constant and emission intensity stems from permanent damages induced by the intense ionizations. In PVT, it is known that ion irradiation drastically changes the chemical structure, leading to hydrogen and hydrocarbon groups desorption before transformation to hydrogenated amorphous carbon at the highest doses, while luminescence is quenched [17]. The previous investigations on IBIL of BC-400 [10-13], non time-resolved and mostly using light ions, evidenced that theses modifications are associated to a degradation of scintillation intensity. Intensity deterioration can basically proceed from two main types of quenching, namely dynamic and static quenching [18,19]. Dynamic quenching results from collisions between excited fluorophores and quenchers that facilitate non-radiative transitions to the ground state. Here, quenchers are species related to the radiation damaging of the molecules of the PVT matrix or the embedded fluorophores [20]. Therefore their concentration have to be an increasing function of the ion fluence. In the simplest case of dynamic quenching, the evolution of the decay constant is given by the Stern-Volmer equation which writes, when quencher concentration is tentatively assumed to be proportional to ion fluence $\Phi$ [18,19]

$$\tau_0/\tau = 1 + \sigma_d \Phi, \quad (1)$$

where $\sigma_d$ is the dynamic quenching cross section. This equation suits to the observed evolution of the decay constant with $\sigma_d = 2.1 \times 10^{-11}$ cm$^2$ (Fig. 4).

Static quenching is related to the formation of ground state complex between the fluorophore and a quencher or to the existence of an effective quenching sphere that diminishes the fraction of fluorescing molecules. Additional static quenching leaves the fluorescence decay unaffected but enhances intensity degradation. In the sphere of effective quenching model, it is assumed that there is instantaneous quenching when the quencher is located inside a sphere of a given volume around the fluorophore and no quenching when the quencher is outside. Combined with dynamic quenching, the evolution of the intensity with fluence according to this model is given by [18,19]

$$I_0/I = (1 + \sigma_d \Phi) e^{\sigma_s \Phi}, \quad (2)$$

where $\sigma_s$ is the static quenching cross section. The fit of the intensity with this equation where $\sigma_d = 2.1\times10^{-11}$ cm$^2$, as determined above, yields $\sigma_s = 8.2\times10^{-11}$ cm$^2$ (Fig. 4).

Thus, the positive deviation of $I_0/I$ from linearity as a function of fluence, that can also be deduced from previous studies, together with the linear increase of $\tau_0/\tau$, evidenced by our time-resolved instrumentation, indicates that the IBIL degradation in BC-400 follows a combination of dynamic and static quenching mechanisms (Fig. 4). Static quenching is the predominant mechanism as the intensity is decreased by a factor of ~3 at final fluence while the decay constant is decreased by about 20%, predominance also illustrated by the factor 4 difference between static and dynamic cross sections.

A convenient, model-independent, measurement of the radiation hardness of scintillators is the critical fluence $\Phi_{1/2}$, that is the fluence required for the scintillation intensity to be divided by two. The mean value obtained in this study with 27-MeV $^{36}$Ar ions having 2.6 MeV/µm mean electronic stopping power is $\Phi_{1/2} = 6.8\times10^9$ ion/cm$^2$. Fig. 5 compiles the critical fluences as a function of mean electronic stopping power from IBIL degradation studies of BC-400. Despite an important scattering in the data, there is a global decreasing trend. Remarkably, there is an internal consistency in the studies of Torrisi [10] and Quaranta et al. [11], both showing a monotonous decrease of $\Phi_{1/2}$ as a function of $<(dE/dx)_e>$. The effects of ion irradiation on polymers were extensively studied last decades and reported in several review or general articles giving several evidences of increasing damage when increasing the ion mass [21-25]. More specifically, under low $<(dE/dx)_e>$ irradiations, aromatic polymers show a very good radiation resistance, orders of magnitude higher than those of aliphatic compounds. However, it has been shown that in several aromatic polymers (polystyrene [26,27], polycarbonate [28], polyethylene terephthalate [29] and poly(p-phenylene sulphide) [30,31]) that the disruption rate of chemical bonds is strongly enhanced when increasing the ion stopping power. Such effect is hence expected in BC-400 which has an aromatic polymer matrix. Our study, performed at the highest mean electronic stopping power, at the same time yields the smallest critical fluence, in line with this trend. Our critical fluence is 3-4 orders of magnitude lower than those for particles with $\langle(dE/dx)_e\rangle <100$ keV/µm and 1-2 orders of magnitude lower than those for particles with $\langle(dE/dx)_e\rangle <1$ MeV/µm. The fit including our and previous data yields $\Phi_{1/2} \propto \langle(dE/dx)_e\rangle^{-1.7}$, i.e. that the critical fluence has roughly a quadratic dependence on the reciprocal of mean electronic stopping power (Fig. 5). This compares well, for instance, with the quadratic dependence of damage cross sections on stopping power in poly(p-phenylene sulfide) (PPS), which is independent on the chemical nature of the bonds [26]. This non-linearity implies that the dose from which the IBIL of BC-400 starts deteriorating is variable and depends on the stopping power of the particles. Therefore, the radiation hardness of plastic scintillators can be strongly weakened under high-density excitations with heavy ions.

**Conclusion**

Our new time-resolved instrument designed to study the dynamics of luminescence under high density excitation was successfully applied to the investigation of BC-400 plastic scintillator under 27-Mev Ar ion irradiation. The apparent decay time constant at initial fluence is similar to that reported in literature. Analysis of the rise time suggests that the total time resolution of the instrument is slightly below one nanosecond. Following the scintillator dynamics with increasing ion fluence, we evidenced a strong degradation of scintillator intensity due to radiation damages. Additionally, due to the time-resolving capability of our device, we evidenced that this is accompanied by a decrease in the decay constant. The combination of both phenomena indicates that both static and, to a lesser extent, dynamic quenching are occurring. Scintillation intensity reaches half of its initial value at a critical fluence of about $10^{10}$ ion/cm$^2$. This value is up to several orders of magnitude smaller than those reported in previous studies, mostly performed using light ions with smaller stopping powers. The compilation of our data with those from the previous studies reveals an almost quadratic dependence of the weakening of radiation hardness on the stopping power of the projectiles. This illustrates the potential limitations of the use of plastic scintillators in heavy ion environments.

**Acknowledgements**

We would like to acknowledge financial support by the European Community as an Integrating Activity 'Support of Public and Industrial Research Using Ion Beam Technology (SPIRIT)' under EC contract no. 227012 as well as by the ERC grant no. 279790. We also would like to acknowledge financial support from the French national accelerator network dedicated to material irradiation EMIR. Experiments were performed at the IRRSUD beam line of the GANIL facility in Caen, France.

**Figures**

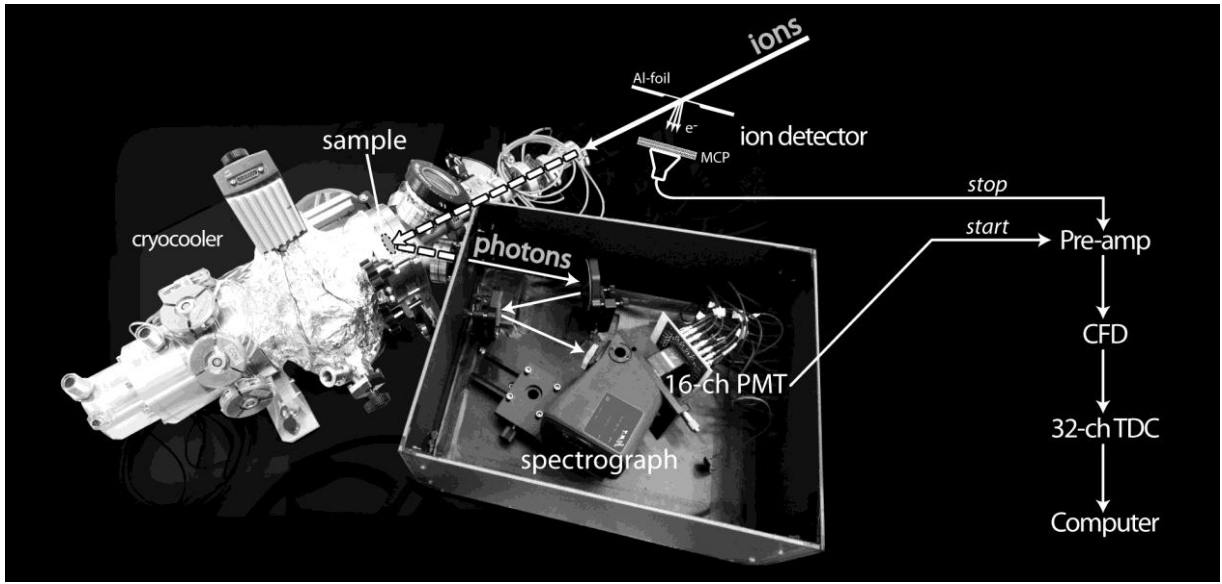

Fig. 1. SPORT instrumental set-up. Note that the second optical table, which extends the wavelength range with 16 additional channels, is absent on the picture since not used in this study.

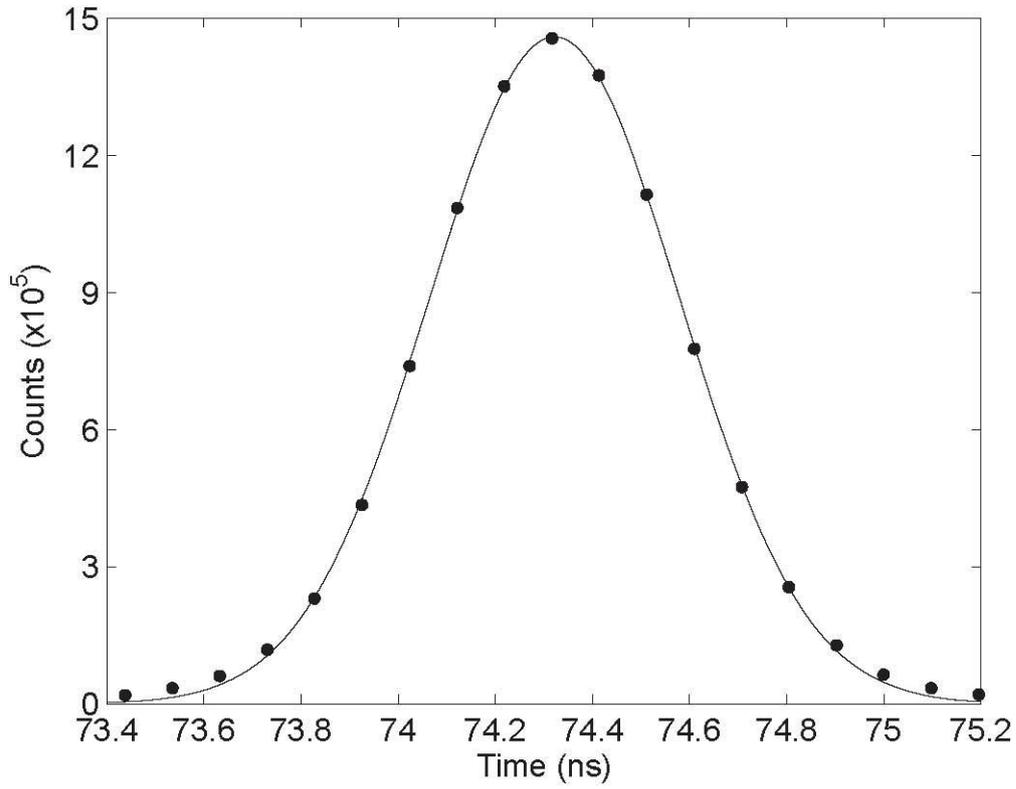

Fig. 2 Ion self-correlation curve, representing the time spread between ions from successive packages of the cyclotron. The solid line represents the Gaussian fit with 0.6 ns FWHM and centred at about 74.3 ns. Note that this latter value corresponds to the repetition rate of the cyclotron (13.45 MHz).

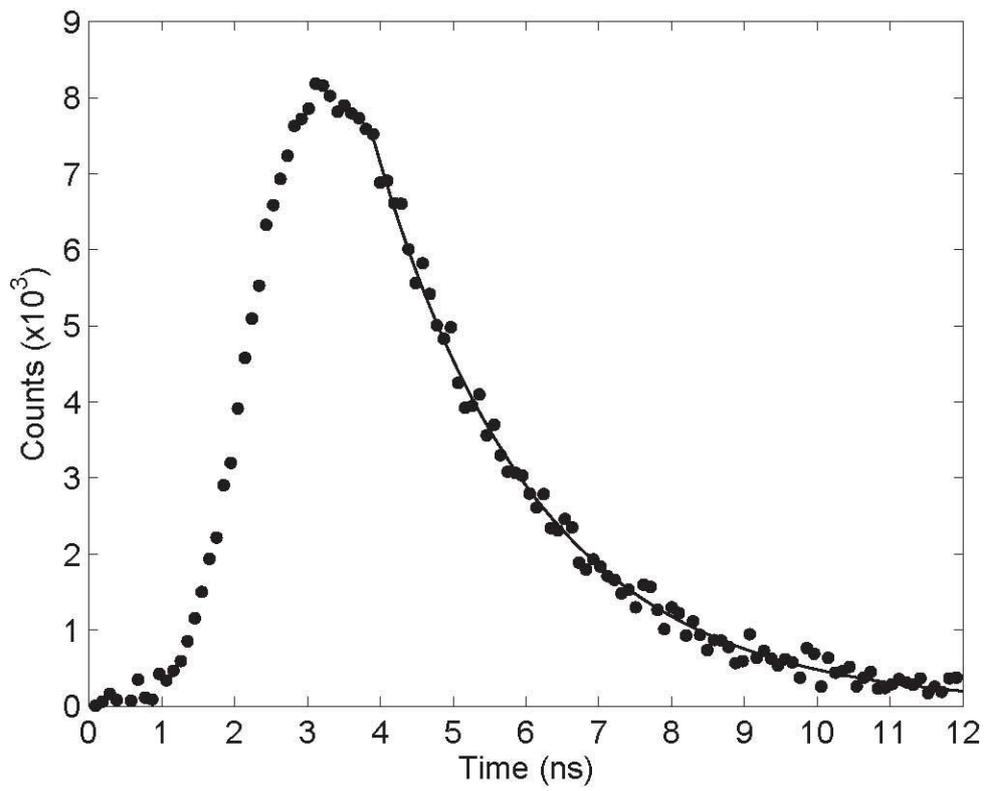

Fig. 3 Emission curve of the 423-nm band of BC-400 with 27-MeV $^{36}$Ar irradiation. The solid line represents the fit of decay with a single exponential.

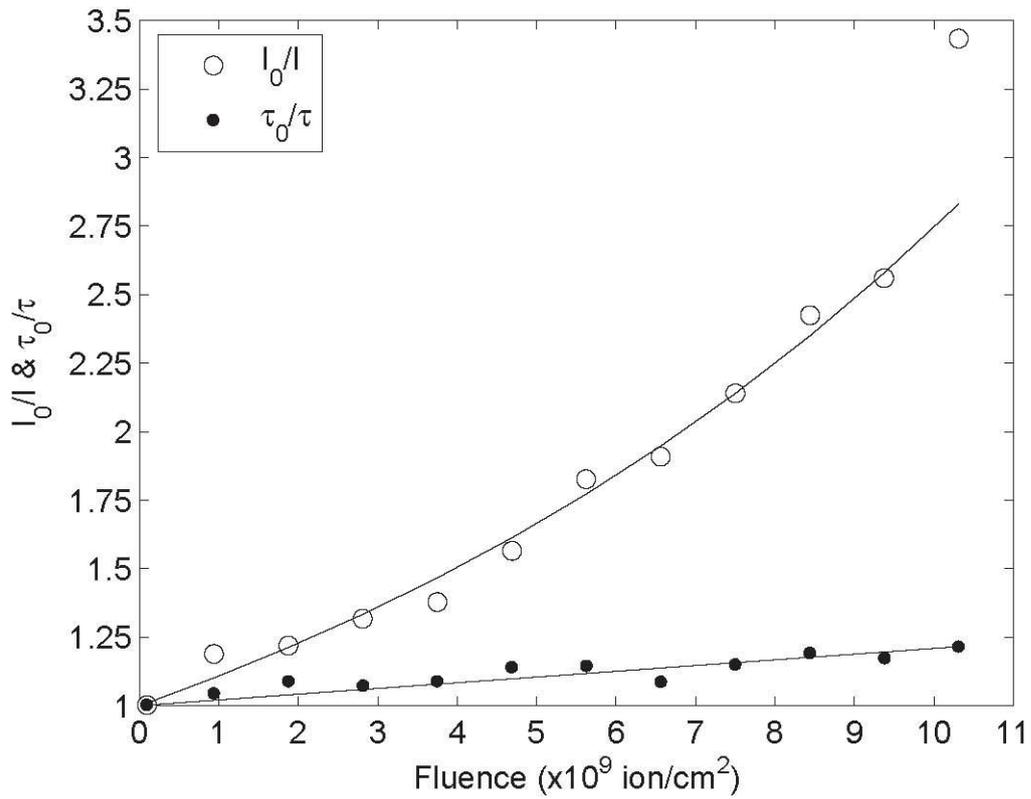

Fig. 4 Stern-Volmer plots of $I_0/I$ and $\tau_0/\tau$ as a function of fluence. Solid lines represents fits of $I_0/I$ and $\tau_0/\tau$ with Eqs. 1 and 2 respectively.

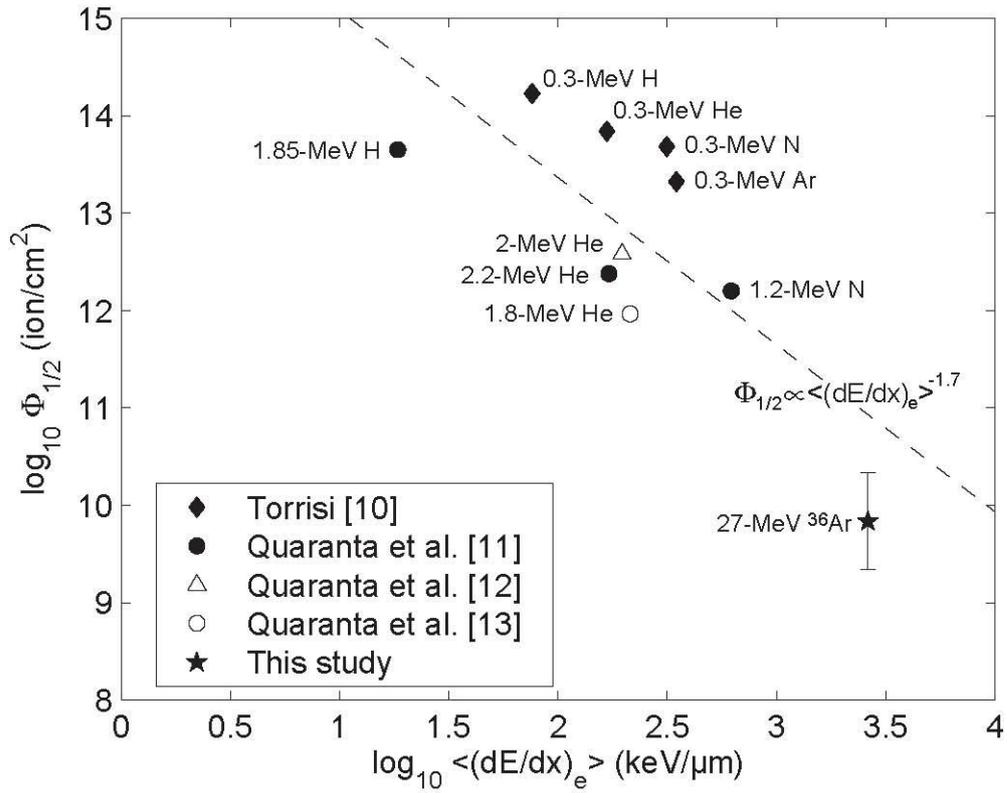

Fig. 5 Radiation hardness of BC-400, illustrated by the critical fluence $\Phi_{1/2}$ at which scintillation intensity is divided by two, as a function of mean electronic stopping power $\langle(dE/dx)_e\rangle$.